\documentclass{article}
\def\wit{\hspace{1pt}}

\def\csup#1#2{\mbox{$#1\hspace{-9pt}\hspace{-#2pt}^{\mbox{
                \tiny $^{\smallfrown}$}}$}}
\def\spil#1#2{\mbox{$#1\hspace{-4pt}\hspace{-#2pt}^{\mbox{\footnotesize '}}$\wit}}
\def\spir#1#2{\mbox{$#1\hspace{-4pt}\hspace{-#2pt}^{\mbox{\footnotesize `}}$\wit}}
\def\acca#1#2{\mbox{$#1\hspace{-4pt}\hspace{-#2pt}^{\mbox{\tiny
$^{\prime}$}}$\wit}}
\def\accg#1#2{\mbox{$#1\hspace{-4pt}\hspace{-#2pt}^{\mbox{\tiny
$^{\backprime}$}}$\wit}}
\def\spilacca#1#2{\mbox{$#1\hspace{-8.5pt}\hspace{-#2pt}^{\mbox{
            \footnotesize '\tiny $^{\prime}$}}$}}

\def\spiraccg#1#2{\mbox{$#1\hspace{-8.5pt}\hspace{-#2pt}^{\mbox{
            \footnotesize `\tiny $^{\backprime}$}}$}}
\def\csupspil#1#2{\mbox{$#1\hspace{-4pt}\hspace{-#2pt}^{\csup{
            \mbox{\footnotesize '}}{-2}}$}}

\usepackage[]{amsmath,amscd,amsfonts,amsthm,amssymb}

\title{Early Greek Thought and Perspectives for the Interpretation of Quantum Mechanics:
Preliminaries to an Ontological Approach.}
\author{Karin Verelst and Bob Coecke}
\date{CLEA-FUND, Free University of Brussels\\ Pleinlaan 2, B-1050, Brussels, Belgium\\
kverelst@vub.ac.be; bocoecke@vub.ac.be}

\begin{document}

\maketitle 

\begin{abstract}
It will be shown in this article that an ontological approach for some problems related to
the interpretation of Quantum Mechanics (QM) could emerge from a
re-evaluation of the main paradox of early Greek thought: the paradox of Being and
non-Being, and the solutions presented to it by Plato and Aristotle. More well known are
the derivative paradoxes of Zeno: the paradox of motion and the paradox of the One and the
Many. They stem from what was perceived by classical philosophy to be {\it the} fundamental
enigma for thinking about the world: the seemingly contradictory results that followed from
the co-incidence of being and non-being in the world of change and motion as we experience
it, and the experience of absolute existence here and now. The most clear expression of both
stances can be found, again following classical thought, in the thinking of Heraclitus of
Ephesus and Parmenides of Elea. The problem put forward by these paradoxes reduces for both
Plato and Aristotle to the possibility of the existence of stable objects as a necessary
condition for knowledge. Hence the primarily {\it ontological} nature of the solutions they
proposed: Plato's Theory of Forms and Aristotle's metaphysics and logic. Plato's and
Aristotle's systems are argued here to do on the ontological level essentially the same: to
introduce stability in the world by introducing the notion of a separable, stable object, for
which a `principle of contradiction' is valid: an object cannot be and not-be at the same
place at the same time. So it {\it becomes possible} to forbid contradiction on an
epistemological level, and thus to guarantee the certainty of knowledge that seemed to be
threatened before. After leaving Aristotelian metaphysics, early modern science had to cope
with these problems: it did so by introducing ``space" as the seat of stability, and ``time"
as the theater of motion. But the ontological structure present in this solution remained
the same. Therefore the fundamental notion `separable system', related to the notions
observation and measurement, themselves related to the modern concepts of space and time,
appears to be intrinsically problematic, because it is inextricably connected to classical
logic on the ontological level. We see therefore the problems dealt with by quantum logic not
as merely formal, and the problem of `non-locality' as related to it, indicating the need to
re-think the notions {\it system}, {\it entity}, as well as the implications of the operation
`measurement', which is seen here as an application of classical logic (including its
ontological consequences) on the material world.
\end{abstract} 

\bibliographystyle{alpha} 

\pagenumbering{arabic} 
\pagestyle{plain}

\section{Introduction}

At the origin of our approach lay two encounters between Greek thought and Quantum Mechanics,
one of them deliberately conceived by its author, the other being a meeting between new
QM-concepts and one of the oldest problems of human thinking about the world. The first
encounter has been presented in a short lecture by C. Piron (\cite{pir76a}, p. 169) in which
he attempted to develop a realistic QM-interpretation based on two concepts fundamental to
Aristotelian metaphysics, viz. potentiality and actuality. The second one is the doctoral
dissertation of D. Aerts \cite{aer81}. It both by content and title dealt with the problem of
the One and the Many, the central theme Plato inherited from fifth century philosopher
Parmenides of Elea. In what follows we will attempt to make clear that these encounters are
not of a purely coincidential nature. We intend to develop in this paper an analysis and
re-evaluation of these old questions and their solutions. Our contention is that this
analysis might amount into new perspectives on the interpretation of QM, since the enigmas
and paradoxes of early Greek thought, and the solutions presented to them in the ``classical
period" are, more than we realize, bound to mark our way of {\it looking at } and {\it
reasoning about } the world\footnote{We find support for the relevance of this position in
Schr\"{o}dinger's fine little book dedicated to the subject (\cite{sch96}, p. 3 sq, p. 159)}.
The thought-instruments developed by Plato and Aristotle, in order to solve the riddles
following out of so called ``pre-Socratic" thought, which are epitomized in the, according to
both classical thinkers apparent, contradiction between the ``worldviews" of Heraclitus of
Ephesus and Parmenides of Elea, are in use up to the present, be it in slightly modified
forms\footnote{``Two great warring traditions regarding consistency originated in the days of
the Presocratics at the very dawn of philosophy. The one, going back to Heraclitus, insists
that the world {\it is not} a consistent system and that, accordingly, coherent knowledge of
it cannot be attained by man. (...) The second tradition, going back to Parmenides, holds
that the world {\it is} a consistent system and that knowledge of it must correspondingly be
coherent as well, so that all contradictions must be eschewed." (\cite{res80}, introduction.)
A clearer \vspace{-0.5mm} contemporary formulation of the classical position can hardly be
imagined!}. Our position will be that this classical contradiction has slipped through the ages
unimpaired, but in different forms, such as to make it hardly recognisable in our present
epistemological and ontological concepts, both in philosophy and in science. A revelation of
this implicit presence by reconstructing the outlines of its historical pathway then becomes
the necessary first step towards an approach for the tackling of problems it eventually
causes in to-day's science. The argument will lead us to the conclusion that the paradoxes
appearing in QM represent such a problem. A sketch of some possible strategies will
complete this attempt at clarification.  Methodologically therefore the arguments in this
paper will be based both on scientific and philosophical grounds; even more, it can be
considered our aim to show that, when it comes to a proper understanding of the significance
and implications of basic ``scientific'' findings, both are inextricably intertwined, since
science, seemingly so different, follows a path of deep conceptions laid down much earlier in
the development of human consciousness. We are tempted to see the origins of the QM-paradoxes
as consequences of the ontological ``choices" of Plato and Aristotle. Their effort concerned
the stabilisation of the world of constant change, thus saving the possibility of certain
knowledge in order to escape the contradictions between stable and unstable, knowable and
unknowable that appear on the level of what happens in reality, as expressed mainly by
Heraclitus and Parmenides. This led to the conception of {\it logic} as a standardising rule
for thinking and, much later, {\it experiment} as a standardising rule for experiencing
(\cite{dij96}, IV: 99, p. 381). Both originated out of needs felt in the context of the
macroworld, and reach now their limitations in the study of the microworld. At the moment
when the interactions between things become as important as the things themselves, the
separating intervention in reality, first conceptualised by Plato and Aristotle, seems to
reveal itself as an illusion. For this reason in this paper no position will be taken in the
debate between rationalists and empiricists. From our point of view these philosophical
stances come down to the same on the level of the ontological question: {\it all} imply the
metaphysical world-structure put forward by Plato and Aristotle as a solution for the
ontological paradoxes raised by pre-Socratic thought.

\section{The historical context} 

Pre-Socratic philosophy (\cite{mou93}, Introduction) is the general name given to a rather
differentiated group of Greek thinkers, living from the 7th to the 5th century BC. The two
main currents that are of concern to us here are the Heraclitean and the Eleatic school
of thought. The Heracliteans claimed to be the followers
of {\bf Heraclitus of Ephesus}, of whose work only some fragments survived. Plato, followed by
Aristotle, assigned to him the {\it flux-theory}, the doctrine of permanent motion and
unstability in the world. The consequences of this doctrine are, as both Plato and Aristotle
stressed repeatedly, the impossibility to develop stable, certain knowledge about the world,
for an object, changing each instant, does not allow for even to be named with certainty,
let alone to be ``known'', i.e., assigned fixed, {\it objective} characteristics. A fragment
that might have inspired Plato (\cite{fow67}, {\it Cratylus}, 66, 402(a); {\it Theaetetus},
72, 160 (d,e)) for this interpretation is the famous 
{\it riverfragment}\footnote{$\pi o\tau\alpha\mu o\csup{\iota}{-1}\varsigma \
\tau o\csup{\iota}{-1}\varsigma \
\alpha\spil{\upsilon}{-1}\tau o\csup{\iota}{-1}\varsigma \
\spil{\epsilon}{-1}\mu\beta\alpha\acca{\iota}{-1}\nu o\mu\acca{\epsilon}{-1}\nu
\ \tau\epsilon \ \kappa\alpha\acca{\iota}{-1} \ o\spil{\upsilon}{-1}\kappa \
\spil{\epsilon}{-1}\mu\beta\alpha\acca{\iota}{-1}\nu o\mu\acca{\epsilon}{-1}\nu
\ \epsilon\spilacca{\iota}{-1}\mu\epsilon\nu \
\tau\epsilon \ \kappa\alpha\acca{\iota}{-1} \ o\spil{\upsilon}{-1}\kappa \
\epsilon\spilacca{\iota}{-1}\mu\epsilon\nu$. {\it In dieselben Fl\"{u}sse
steigen wir und steigen wir nicht, wir sind und wir sind nicht} (\cite{die71}, DK 49a, p. 161).} in
which Heraclitus states that because of the ever ongoing flow of the waters, it is impossible to
step into the {\it same} river twice. He stresses the ever ongoing change or motion (both being
aspects of the same process) that characterises this world and its phenomena. This is so because
Being, over a lapse of time, has no stability. Everything that {it is} at this moment changes at
the same time, therefore {it is not}. This coming together of Being and non-Being at one instant is
known as {\bf the principle of coincidence of opposites}\footnote{The formula stems from Nicolaus
Cusanus, a thinker on the edge of the Middle Ages and early Renaissance. But compare e.g. Cusanus's
``circular theology" with the meaning of the ``Ouroboros'', the circular snake of ancient alchemy.
For Cusanus, see (\cite{dij96}, p. 248 sq.); for the \vspace{-1mm} Ouroboros, see (\cite{luc85},
p. 367-368).}. It is crucial to see that this principle is connected to the possibilty of motion,
for being in motion implies to be and to be not at the same time at a certain place on a certain
moment. It further implies the unity of the world in the sense that there are no separated
objects\footnote{$o\spil{\upsilon}{-1}\kappa \ \spil{\epsilon}{-1}\mu
o\csup{\upsilon}{-1} \ \spil{\alpha}{-1}\lambda\lambda\accg{\alpha}{-1} \ \tau o\csup{\upsilon}{-1}
\lambda\acca{o}{-1}\gamma o\upsilon \ \spil{\alpha}{-1}\kappa
o\acca{\upsilon}{-1}\sigma\alpha\nu\tau\alpha\varsigma\ \ \spir{o}{-1}\mu o\lambda
o\gamma\epsilon\csup{\iota}{-1}\nu \ \sigma o\phi\acca{o}{-1}\nu \ \
\spilacca{\epsilon}{-2}\sigma\tau\iota\nu \ \spiraccg{\epsilon}{-1}\nu \
\pi\acca{\alpha}{0}\nu\tau\alpha \ \epsilon\csupspil{\iota}{0}\nu\alpha\iota$. Haben sie nicht mich,
sondern den Sinn vernommen, so ist es weise, dem Sinne gem\"{a}\ss\  zu sagen, alles sei eins.
(\cite{die71}, DK 50, p. 161).}, its ontology is {\it dynamic}. However, this does not mean that
Heraclitus meant to say that there is absolutely no stability in things. The Heraclitean concept of
harmony\footnote{On the (ir)relevance of the flux-theory for Heraclitus's thought and possible
interpretations of his concept of
$\spir{\alpha}{0}\rho\mu o\nu\acca{\iota}{0}\alpha$, see (\cite{kir75}, introduction); also
(\cite{kir83}, p. 193).}, closely connected to the principle of coincidence of opposites, rather
points in the direction of a permanent and momentaneous re-instantiation of things in the world in
a web of totally interconnected events. Reality is One. This does not imply the ``unreality" of the
things we experience, it states our absolute interconnectedness with the world we experience. 
Tradition transmitted as its counterpart the Eleatic school, named after its inspirator {\bf
Parmenides of Elea}. Of him more extant textfragments are known, although there remains doubt about
their true meaning. This perception of tradition started with Plato and Aristotle, who saw him as
the opponent of Heraclitus, learning the non-existence of motion and change in reality, reality
being absolutely One, and being absolutely Being\footnote{$\tau\accg{o}{0} \
\spil{\epsilon}{0}\accg{o}{-1}\nu \ \spilacca{\epsilon}{-2}\sigma\tau\iota$ {\it Being is}.
(actually it says: {\it The Now-Being is}). (\cite{rie70}, p. 45).}. This brings to mind the
paradoxes of Zeno of Elea (\cite{kir83}, p. 263 sq.), according to tradition a disciple of
Parmenides (\cite{fow67}, {\it Parmenides}, 204-206, 128(c-d)). The most famous of them is the
paradox of motion, a less famous one deals with the possibility of there being separate things in
the world: the paradox of the One and the Many. Zeno's until now essentially unresolved
paradoxes\footnote{Dijksterhuis remarks sharply: ``(...) we don't know how gravity manages to give
velocity to the body, and very often the given explanation comes down to pretending to understand
on the microlevel what should be explained on the macrolevel. But we know since Zeno of Elea, that
here lurks an essential difficulty of the conceptualisation of motion... " (\cite{dij96}, p. 203,
our translation). See also the comments on modern attempts to explain Newton's laws of gravitation
by means of different kinds of bizarre particles, the latest one being the theory of the graviton,
by C. Piron in the text of a conference given November 8, 1996 for CLEA in Brussels (\cite{pir96},
p. 2).} were intended to show that motion and separated things cannot exist, precisely because
their existence demands the coming together of being and non-being at the same time on a certain
place at a certain moment in a world whose fundamental feature it is to exist. Non-Being is a
contradiction in itself (\cite{vla93}, p. 193). The atomists' reaction to the Eleatic enigma
(\cite{kir83}, pp. 407-408) {\it and} to our experience of change was the introduction of non-Being
in the form of the ``void''. The void is nothing else than non-Being, in which Being, in the form
of a plurality of Eleatic Ones, moves about\footnote{$\tau\acca{o}{-1} \
\kappa\epsilon\nu\acca{o}{-1.5}\nu$ ``void'' literally means non-being. That you can grant for the
possibility of motion by the introduction of the void was discovered \vspace{-0.5mm}Melissus, a
follower of Parmenides, and of whom Leucippus, founder of atomism, was a disciple according to
tradition (\cite{kir83}, pp. 397-398). Furthermore $\spilacca{\alpha}{-1}\tau o\mu o\varsigma$,
i.e., without
$\tau o\mu\acca{\eta}{-1}$, without cuts, unseparated in itself, undivided. It is important to see
that this originally meant that there are no ``parts" in this Being (in this entity, again
literally the same word), between which there can be ``parts" of non-Being, the non-Being having a
different character than the all-pervading, homogeneous void or empty space \vspace{-1mm} of later
times.}. They can stick together but not mingle with each other, because of their Eleatic
nature\footnote{$\kappa\iota\nu o\csup{\upsilon}{-1}\nu\tau\alpha\iota \ \tau\epsilon
\ \sigma\upsilon\nu\epsilon\chi\csup{\omega}{-1}\varsigma \ \alpha\spir{\iota}{-1} \
\spilacca{\alpha}{-1}\tau o\mu o\iota \ \tau\accg{o}{-1}\nu \
\alpha\spil{\iota}{-1}\csup{\omega}{-1}\nu\alpha \ {\it (...)} \ \spilacca{\eta}{-1} \
\tau\epsilon \ \gamma\accg{\alpha}{-1}\rho \ \tau o\csup{\upsilon}{-1}\
\ \kappa\epsilon\nu o\csup{\upsilon}{-1} \ \eta \ \phi\acca{\upsilon}{-1}\sigma\iota\varsigma \
\spir{\eta}{-1} \ \delta\iota o\rho\acca{\iota}{-1}\zeta o\upsilon\sigma\alpha \
\spir{\epsilon}{-1}\kappa\acca{\alpha}{-1}\sigma\tau\eta\nu \
\alpha\spil{\upsilon}{-1}\tau\accg{\eta}{-1}\nu \ \tau o\csup{\upsilon}{-1}\tau o\nu \\
\pi\alpha\rho\alpha\sigma\kappa\epsilon\upsilon\acca{\alpha}{-1}\zeta\epsilon\iota$ {\it Les atomes
se meuvent contin\^{u}ment durant l'\'{e}ternit\'{e} (...) Car la nature du vide, qui s\'{e}pare
chaque atome en lui-m\^{e}me, produit cette effet (...)}. (\cite{con87}, in the `lettre \`{a}
H\'{e}rodote', pp. 102-103)}. {\it Change thus reduces to motion}, but the origin of this motion,
and the paradoxical acceptance of the being of non-Being, remain riddles unsolved. Because - from
the Eleatic point of view - in the real world everything {\it is}, eternally and indivisibly, it is
impossible to speak of something that is not in a way that makes sense.  But, as said before, the
``contradiction" seen by classical philosophy between Heraclitus and Parmenides is not necessarily
a correct understanding of the earlier ``philosophies". One could as well infer that Heraclitus and
Parmenides do articulate the {\it same} world-experience, the former as the experience of reality
over a lapse of time, the latter as the experience of the absolute reality of {\it this moment} (to
understand better what this means, try to deny by yourself you are {\it experiencing} yourself as
existing {\it at this moment})\footnote{Mediaeval thought knew this experiencing of experience
as the {\it nunc stans}, the ``standing now". (\cite{are78}, p. 210)}. This has nothing to do
with the intellectual question what it {\it means} to exist, or whether our existence is
``real" or not. These questions concern things ``as such", objects, and their identity in
past and future. But this type of interpretation - which is the interpretation of classical
philosophy {\it and} of science, and which entails a {\it representation} of reality outside
of its actual and momentaneous experience doesn't make sense, because for Heraclitus no
things ``as such" do exist, and for Parmenides there is no motion, which implies that there
is no time. It is our conviction that, rather than revealing the contradiction between the
``thought-systems" of the two pre-Socratic ``philosophers", Plato's interpretation reveals
the difference between {\it their} world-experience and what we {\it think} to be ours,
constructed on the rational base laid down by classical philosophy. The non-existence of
metaphysical worldviews in the pre-Socratic period is then due to a different kind of
awareness of one's being-in-the-world that characterized the transition from mythical
awareness to rational self-consciousnes (\cite{jay76}, pp. 67. sq.). The hallmark of this
awareness is transparance for the stream of events that constitutes the world (\cite{cam91},
pp. 81 sq.); there is no such thing as the separation between subject and
object\footnote{``it is still the primary function of the {\it noos} to be in direct touch
with ultimate reality." (\cite{fri93}, p. 52)}. This separation precisely coincides with the
coming-to-be of {\it rational} self-consciousness. In philosophy, this change will be
codified in the metaphysical systems of Plato and Aristotle. The process found its completion
only in the early modern period, when rationalised self-consciousness developed the
scientific way of observation and explanation of the world. The problem of change (and motion
as a special case of change) and the problem of the existence of separable objects in the
world appear as two sides of the same coin. It is at the origin of {\bf the principle of
contradiction}, formulated by both Plato and Aristotle.   
 
The contradicting conclusions deriving from pre-Socratic philosophy were of a major concern
to Plato and Aristotle, because they challenged the existence of truth and certainty about
the world and therefore about the actions of human beings in it. This uncertainty had given
rise to a philosophical discipline, Sophism, that simply denied any relation between reality
and what we say about it (\cite{fow67}, {\it Theaetetus}, 42, 152(d,e)). Its subjectivism
stems from a radical empiricism, which holds that things are for me as I perceive them. But
since reality as we perceive it is always in a process of permanent change this implies, as
Plato points out in the Theaetetus\footnote{Plato explicitly refers to Protagoras's ``man is
the measure of all things, existing and non-existing". (\cite{fow67}, {\it Theaetetus},
160(d,e))}, also the non-existence of stable, individual things in the world. But then again
this means that Sophism is nothing else but an instance of the ontology of permanent change,
already formulated by earlier thinkers like Heraclitus. This is what, according to Plato,
follows for knowledge out of the ontology of universal change, and {\bf ``everything is
equally true"} reveals itself as {\bf the epistemological formulation of the coincidence of
opposites}, which we met before as the base of the universal change theory. On this soil the
principle of contradiction rests, because if you {\it allow} contradiction, you will be
allowed to say whatever: {\it ex falso quodlibet} (\cite{fow67}, {\it Theaetetus}, 150,
182(e)-183(a)). But how to conceal this prohibition with our experience of permanent change
in the world? This will only be possible by stabilising human world-experience in a
world-{\it picture}, strong enough to survive the paradoxical present into the past and the
future\footnote{The necessary condition that made possible this construction of stabilising
world-pictures or ``worldviews", was the earlier coming-to-be of the ``inner mind-space", in
which the non-present could be re-presented as present (\cite{jay76}, p. 54 sq.).}. Strong
foundations must be laid to grant the possibility to experience entities as objects outside
of the stream of events, and therefore to speak about them in an ``objective" way. That is
where, from our point of view, the true origin of philosophy and in a later stadium science
are to be situated.  The main problem for Plato and Aristotle can be described therefore as
the construction of a {\it world-picture} that would 1) grant the ultimate stability of
things, neccesary as a solid base for certain knowledge, and at the same time 2) allow for
non-being, necessary for the existence of the change, motion, manifold etc., that we
experience by our senses. The common feature of both their solutions was {\bf the division
of the world in two separated, though connected, layers:} an unchangeable, motionless
`Parmenidean' or `Eleatic' one which grants certainty about both objects and names, and a
second `Heraclitean' one, changeable and moving, which allows change and motion in the world
as presented to our senses. This feature of a two-layered world (a ``world behind the
world"), is what makes their worldviews {\bf metaphysical} (\cite{are78}, {\it One /
Thinking}, p. 23). In respect to this, we consider the differences between the two as rather
superficial. The ways they choosed to achieve this are nevertheless very different. They will be
treated in more detail below.

\section{The solutions of Plato and Aristotle}

{\bf Plato's} system displays an explicit two-world structure: the eternal world of Forms or
Ideas, and the world of changeable phenomena. The relation between the two consists in the
participation of the phenomena in the Forms, the Forms granting them ultimate stability and
knowability. It is the hierarchy in the level of Being (in the level of reality) of the
Forms and the participation of things in the Forms (the {\it participation theory}) that
allows for the possibility of an object to have contrary properties without creating a
contradiction between them that would arise out of the simultaneous presence of Being and
non-Being\footnote{Plato's elaboration for the Forms Being, Motion and Rest in
(\cite{fow67}, {\it Sophist}, 388, 250(b,c)). A more general ontological formulation can be
found e.g. (\cite{fow67},{\it id.}, 392, 251(d); 413, 256(d,e)).} The reasoning runs as
follows: if there be a Form ``Being" and another Form ``motion", then it is clear that
``Being" must have a deeper, broader level of reality, because everything that moves exists,
but not everything that exists moves. Something can be ``non-moving" in the sense that it
{\it is} because it participates in the Form ``Being", and {\it is not} in the sense that it
participates in the Form ``Rest" and therefore doesn't move. Because of {\it the different
level of reality} of the Forms ``Being" and ``Rest", this will not lead to the ontological
impossibility of something {\it not being} on the level of Being itself. This constitutes
the ontological part of his system. On the level of knowledge this is reflected in the
structure of language. Plato is the first to discriminate between the predicative and
existential use of the verb `to be'\footnote{``Platonists who doubt that they are spectators
of Being must settle for the knowledge that they are investigators of the verb ``to be".
(\cite{owe71}, p. 223)}. This is possible because classes of concepts do ``mingle" the same
way as classes of Forms (\cite{fow67}, {\it Sophist}, 400, 253(b,c); id., 401, 253(d,e)). On
the epistemological level, this discrimination plays exactly the same role as the
existence-hierarchy of the Forms does ontologically. It allows for the possibility to {\it
speak} about a thing having contrary properties without ending up in contradiction. Plato
thus is the first to formulate the {\bf principle of contradiction}\footnote{Already in the
Phaedo: (\cite{fow67}, {\it Phaedo}, 348, 101(d,e)).}. However, since his epistemology
remains fully embedded in his ontology, the principle {\it follows as a property of the way
we can speak about the world directly out of the participation theory}. The principle of
contradiction is {\bf the formulation on an epistemological level of the participation
theory}\footnote{Epistemological formulation: \cite{fow67}, {\it Sophist}, 414, 257(a,b,c);
{\it id.} 418, 258(b,c).}. Plato's epistemology contains, because of this principle, and
therefore {\it because of the existence of the Forms}, an implicit logical
structure\cite{cas73}. The fundamental (and unproven) axiom on which his system rests is the
existence of the Forms proper. The stability of things and their knowability is granted by
an essence (the Forms) existing before and apart from them. The problem of motion gets its
solution by means of the {\bf degrees of reality} that exist between the Forms mutually. In
this way he gives a foundation to the stability of the world {\it and} to its knowability,
{\it without excluding properties like change  and motion out of it}. Thus he escapes the
pre-Socratic enigma.   

{\bf Aristotle} solves the problem of stability and knowledge - of stability as a necessary
condition for knowledge - in an at first glance totally different manner. But he starts from
{\it the same premiss:} that the world, experienced as external, should be knowable as such,
knowable objectively - i.e., in its quality as a collection of objects - by a subject. As
said before, this separation of the world and the knower causes the falling apart of the
world in the stable, knowable, Eleatic {\it noumenon} (essence), and the unstable
Heraclitean {\it phenomenon}, thus yielding a representation, {\bf a metaphysical
``worldview"} problematic {\it vis-\`{a}-vis} the changing reality open to experience. The
reasons for the Stagiryte's rejection of Plato's system are the difficulties that are raised
- from his point of view - by the Form-ontology. The difficulties are threefold: the {\bf
Third Man, the unlimited number of Forms}, and {\bf motion} (\cite{tre96}, I, 990b(15) -
991a(8); there also footnote c, and 991a(9 sq.)). Aristotle's objections reduce to one major
theme: {\it the rejection of the Forms as stabilising essentials existing separatedly from
the things they instantiate}. How then does he guarantee the existence of stable things that
can undergo change and motion without allowing for the coincidence of opposites? And how
does he save the possibility to speak about them as {\it being} and {\it non-being} without
falling into contradiction? With him, and contrary to Plato, the Eleatic and Heraclitean layers coincide
in one world. A {\it thing} (object of experience) is a {\it essential form}, a {\bf
substance} which realizes itself in an undifferentiated material receptacle\footnote{The
$\spir{\upsilon}{-1}\pi o\kappa\epsilon\iota\mu\acca{\epsilon}{-1}\nu o\nu$. See (\cite{tre96},
VII, 1028b(84)-1029a(34))}, which can be seen as a {\it substratum} for existence, not as existence
itself. Matter in itself has neither individuality, nor quality. {\it A thing consists of Form and
matter at the same time}. Things therefore are not reflections of idealised Forms in a separated
world, but instantiations of Form - termed substance\footnote{It will be
noted that ``substance" in the Aristotelian sense has nothing to do with the material
connotation that seems evident in {\it our} use of that concept. Aristotelian reality on
the level of matter is a continuum; it is the Form or substance that separates Being from
non-Being.} -  in matter. This instantiation or realisation
is seen as a process in a course of functional development that leads to a certain
endpoint, and in this sense as the ``goal" to be achieved. How, then, does he explain and
justify the nature of things in their coming-to-be and being? This he does by introducing
the {\bf theory of the four causes}. These causes are not to be interpreted in our strict
causal sense; they represent the {\it reasons that  make that something is what it is}
(\cite{tre96}, I, 983a(24-34)). They are not {\it causes} but {\it
becauses}\footnote{$\alpha\spilacca{\iota}{-1}\tau\iota o\varsigma$, cause, bears connotations
different from the modern concept of causality. Causal expressions in both the Platonic and
Aristotelian sense would include (apart from the `strictly causal' ones): Why is this statue so
heavy? {\bf Because} it is made of bronze. Why is he taking after-dinner walks? {\bf Because} of
his health. See (\cite{vla71}, p. 134).}. The {\bf material cause} is the undifferentiated
substratum for existence, in which the essential nature of things will find its expression. The {\bf
formal cause} is exactly this essence, the thing's substantial nature. But because this evolution - this {\it motion towards} - takes place in and
through the material substrate, it is a process that can never be completed. The {\bf efficient cause}
represents the influences from the outside world that cause the process of motion towards
realisation of its true nature. The {\bf final cause} is the endpoint of this realisation,
the completion of {\bf the transformation from potentiality to actuality}. {\it Why} does
Aristotle separate potential from realised being? This is the core of his metaphysical
system, because the transformation from potential to actual is his way of understanding {\bf
motion}. An entity that has realised its substance doesn't change or move anymore. But this
will never be the case for a particular thing in the world, for a thing can only be in
motion {\it in reference to something else}. And since an endless regression of causes of
motion would be absurd, he postulates the existence of the truly actualised and therefore
motionless Form, the First and motionless Mover, God. An even more important point is that
the impossibility of the total actualisation of the things in the world follows out of
their being form and matter at the same time: the entity {\it is} not completely its
actualised self, it {\it is} only in reference to something else. The only truly actualised
thing - true Form - is God. {\bf Here we find the ontological ground for the contradiction
principle:} an entity cannot be in the potential {\it and} in the actual state {\it in
reference to one single other thing} (\cite{tre96}, IV, 1009a(24-39)). This allows
Aristotle to unite the Eleatic and Heraclitean `worlds' of his metaphysical system. We
bring to mind that his system was {\it constructed} this way to obtain this result, to
grant the possibility to deal with the apparent paradoxical nature of `real' reality:
stability and motion should both be accounted for. The world, thus stabilised
ontologically, can now be made accessible to thinking. How does he construct a framework
for {\it knowledge} such that a relation to this stabilised, but divided reality can be
achieved? Here we enter the vast area of {\bf Aristotelian logic}. The principle
of contradiction, based on the separation of being and non-being in the world, can
now be established as {\it the basic axiom} for correct thinking (\cite{tre96}, IV,
1005b(8-34)). Although Aristotle states explicitly its {\it unprovability} (\cite{tre96}, IV,
1005b(35)-1006a(16)), its introduction is justified in the framework of his metaphysics,
where the danger that it would cause the emergence of a static, Eleatic world-picture,
incompatible with our experience, had been neutralised. The three fundamental principles of
classical (Aristotelian) logic: the {\bf existence of objects} of knowledge, the {\bf
principle of contradiction} and the {\bf principle of identity}, all correspond to a
fundamental aspect of his ontology. This is exemplified in the three possible usages of the
verb ``to be": existential, predicative, and identical. The Aristotelian syllogism always
starts with the affirmation of existence: something {\it is}\footnote{With Aristotle,
negation always is a secondary step in the process of reasoning.}. The principle of
contradiction then concerns the way one can speak (predicate) validly about this existing
object, i.e. about the true and falsehood of its having properties, not about its being in
existence. The principle of identity states that the entity is identical to itself at any
moment (a=a), thus granting the stability necessary to {\it name} (identify) it. It will be
clear that the principle of contradiction and the principle of identity are closely
interconnected. In any way, change and motion are intrinsically not provided for in this
framework; therefore the ontology underlying the logical system of knowledge is essentially
{\it static}, and requires the introduction of a First Mover with a proper ontological
status beyond the phenomena for whose change and motion he must account for.

These different positions regarding the stable essence of things will cause the {\it Fight
of the Universals}, the question whether the substances precede (ante re) or coincide with
(in re) the things they instantiate. During the Middle Ages this debate will give rise to a
third possible position: nominalism. It holds that substances (Universals) do not exist
except for our mental activity. But then the debate had already shifted in a purely
epistemological direction, while at its origin were mainly ontological questions. That these
questions even in the epistemological treatment of nominalism don't dissappear, becomes
clear when one considers the difficulties each nominalist theory has to grant soundly for
the possibility to use general concepts, an indispensible tool for scientific theory
(\cite{car92}, p. 85). 

\section{Related conceptions in early modernity}

It is well known that the transformation from medieval to modern science coincides with the
abolition of Aristotelian metaphysics as the foundation of knowledge. Not abondoned until
the twentieth century however was Aristotelian logic as a base for reasoning. Our aim until
now can be summarised as {\it showing that the main principle of syllogistic logic, the
principle of contradiction, contains itself an ontological rule}. The rule is that, contrary
to our daily experiences, intellectual processes be standardised to remove change (and as a
special case: motion) out of the world to assure the possibility of naming and classifying
unambiguously entities as objects. But the change and motion we experience in the world do
remain. In the Stagiryte's system, the possibility for change and motion was granted for
exclusively on the ontological level (the transformation of potentiality into actuality).
Furthermore, although Aristotle separated the disciplines of the theory of Being from the
theory of Reasoning (i.e., ontology from epistemology), we showed above that the latter's
basic rules and categories are ontological principles as well. Dropping Aristotelian
metaphysics, while at the same time continuing to use Aristotelian logic as an empty
``reasoning apparatus" implies therefore loosing the possibility to account for change and
motion in whatever description of the world that is based on it. The fact that Aristotelian
logic transformed during the twentieth century into different formal, axiomatic logical
systems used in today's philosophy and science doesn't really matter, because the
fundamental principle, and therefore the fundamental ontology, remained the same
(\cite{eps90}, p. xix). {\bf This ``emptied"}\footnote{This ``emptiness" is different from
the `emptiness of twenthieth century formal logic. But the latter can be seen as a natural
consequence of the former, since it was developed to deal with logical problems that arise
out of the ontological nature of the rules of logic, as we hope to make clear further in
this article. This applies e.g. to Russell's ``theory of types" (and to all theories based
on ``meta-reasoning"), that can be read as modern (i. e., ``purely epistemological")
theories of categories in the Aristotelian sense.} {\bf logic actually contains an Eleatic
ontology}, that allows only for static descriptions of the world. From this point of view,
the debate during the Renaissance between the proponents of Aristotelian natural physics
and the re-emerging corpuscular theories, can be seen as a debate on how Aristotelian logic
as the base of reasoning, given its inherent ontological nature, can be brought into
agreement with the changing world of our senses\footnote{This of course is by no means to
say that the debate was considered this way by those who where implied. A brilliant
exposure of the backgrounds of the debate between Galilei and the jesuit scientists can be
found in
\cite{red83}. On Newton's backgrounds, see (\cite{dob88}, p. 55 sq.). A more general
discussion in \cite{tou84}.}. This viewpoint also implies that the metaphysical structure
fundamental to the older philosophical systems actually remains present in science, be it in
a different and, in fact, less clear way. 

Our aim is not to describe all subtle differences between the alternative conceptions of the
``real world" advanced at the verge of modern science; this has been done by others in a
brilliant way ; e. g. in \cite{dij96} and
\cite{jam93}. Our concern now is to see whether it is possible to bring at the surface the
essential characteristics of a common line of reasoning which would allow to place science
back into the philosophical development scetched above, and to check whether this clarification of its
its fundamental concepts sheds new light on present-day questions related
to the enigma of Quantum Mechanics. Our argument will lead us to the conclusion that this is
the case indeed: the ontological role of the Eleatic and Heraclitean layers in the
metaphysical reconstruction of reality is played in science by the increasingly absolute
conceptions of space and time, instead of but necessarily correlated to the development of
modern conceptions of the nature of ``matter". This becomes manifest in the
``desubstantialisation" of space and in the increasing parallellism between ``space" and
``time". We will briefly consider the role of ``experimentation" as an observational
practice designed to apply the ontological rule present in the scientific way of reasoning
on our world-experience, by changing ``perception" into ``observation". The formalisation of
the empirical component of the cognitive, ``objectified" world-experience made it possible
to bring perception into agreement with the ontological structure of logical reasoning, {\it
not} the other way around. Therefore the ``epistemological" revolution brought about by
science can be described not as the abondoning of metaphysics, but as the complete {\it
absorption} of the metaphysical structure into the procedures of its formalised
``operational" components, cognitive {\it and} empirical, of this ``objective" (i.e.,
objectified) world-experience. The obscured relation in science between ``act" and
``perception" then allows for the conception of logical reasoning as a {\it representation} of
the ontological structure of reality {\it and} for the succesfull application of science to
the natural world: reality is adapted to the ontological structure of science, not {\it vice
versa}\footnote{The Greeks, whose intuitions about the relation man-world seem to have been
often more clear then ours, were sharply aware of the nature of technical interference with
the course of reality. One of the original meanings of the word mhxan®, tool, technical
device, is `trick, deceit'. Whether this justifies their distrustful attitude towards it is
of course a different matter. This ancient clarity also holds for more fundamental concepts.
A quotation from von Fritz is worthwile: ``(...) the concepts of the ``obscure" Heraclitus
are all perfectly clear and can be very exactly defined. In contrast, the empiricist Sextus,
whose arguments seem so clear and easy to many readers, has no clearly identifiable concept
of either {\it logos} (``reason") or {\it nous} (``mind") at all. {\it Nous} with Sextus is
either identified with {\it logos} or considered a manifestation of it. {\it Logos}, where
Sextus speaks in his own name, is most often ``logical reasoning" or the capacity of logical
reasoning (...). But where Sextus reports the views of other philosophers, logos becomes
just the alternative to {\it aisthesis} (``perception"), whatever this alternative may be,
and so loses all clearly identifiable meaning. {\bf Yet it is highly illustrative of the
change which the concept of nous had undergone} between Heraclitus and Sextus that Sextus,
in trying to explain Heraclitus' concept, begins by connecting it with a term the
preponderant meaning of which is ``reasoning" and ends by almost identifying it with
``sensual perception." {\bf Heraclitus' own concept of {\it nous}}, as we have seen, {\bf
was clearly distinguished from both} but somewhat more nearly related to the latter than to
the former. (\cite{fri93}, pp. 42, 43.) Our boldtype and translation of Greek terms.}. The
certain base for knowledge is thus granted for in the most absolute sense, the circle is
closed: ontology is not a ``problem" anymore. In this context the problem of the validity of
knowledge takes on a new shape. Epistemology, the critical commentary of the process of
science, becomes more and more {\it the} philosophical discipline, and can be considered as
a discipline or theory of knowledge not separated from, but ``without" metaphysics: it
nourishes itself ontologically on science. The apparently sole problematic point concerns
the relation of the ``subject" with the objective world, - the debate between empiricists and
idealists, a ``question" that for evident reasons never can be ``solved" inside its
framework: the relation between knowledge and the real world that is implicitly supposed
here is replaced in science by the relation between knowledge and a {\it reconstruction} of
the real world via an ``empirical" procedure containing an ontological rule that shapes the
relation between the human ``observer" and reality on a much deeper level. The separation
between subject and object has by now been completed. Let us now see how early modern
science solved the problem of ``refilling" logic with an ontology that allows for a world of
change and motion, and therefore for the description of the world of our sensual experience. 

\section{The Solutions of Early Modernity}

At this point in our analysis our problem can be summarised as follows: In the course of a
process governed mainly by religious and societal conflicts, early modern natural philosophy
emancipated itself from its Aristotelian metaphysical foundation. Although indispensible for
its correct understanding, these religious and societal influences do not concern us here;
we will confine ourselves to an investigation of {\it how} this emancipation was achieved,
and {\it by which intellectual means} the problems that arose out of it were solved by its
main executors. Therefore we will look into the altenatives that were formulated for the
Aristotelian conceptions that came under attack first, viz. of matter and motion, and how
the reconciliation with the Eleatic ontology, present unimpaired at the core of Aristotelian
logic, was granted for in a way that remains commensurable with the phenomena of change and
motion in the natural world. The first step was taken by {\bf Galilei} in an attempt to
explain at first the behaviour of light, later on of all manifestations of matter that are
accesible to the senses (\cite{red83}, chapter I), by re-introducing Democritian
atomism as theory of matter and of our perception of it. This implies a {\it quantitative}
explanation of the qualitative changes that we experience in the macroworld on the level of
the basic structure of material bodies; change reduces to motion. Following the same
philosophical line, he established the description of the physical behaviour of phenomena
observed in standardised conditions by way of mathematical laws, allowing for future
verification\footnote{Galilei's famous dual method, composed of {\it Metodo risolutivo}
(analytical method based on experimental data) and {\it metodo compositivo} (synthetical
method, generalises the principles found by the former and proves by prediction and
verification that they hold for the phenomena under study). (\cite{dij96}, p. 259.)}. It was
indicated before that atomism requires the introduction of a concept of non-Being or
``void", the precursor of ``empty space", and the indestructibility and incorruptibility of
the ``a-tomoi" if it wants to remain logically consistent. Equipped this way it would
provide the perfect metaphysical stuffing for logic's hidden ontology, if there would not
remain some major deficiencies to deal with, viz. the total lack of understanding of the
{\it origin} of the motion that it makes logically possible, or the question how the
invisible atomoi constitute the sensible things in the world. Nevertheless, in the context
of our argument, Galilei's atomism is of uttermost importance. Not only because it was the
true reason for his condemnation by the Church \cite{red83}, but
because the re-introduction of corpuscular explanations for the nature and properties of
material bodies and the enigma of change and motion was a possible alternative to {\bf the
Cartesian ontology of the identity of matter and space}. This identity followed from
Descartes's contention that {\it extension} and {\it position} are the essential attributes of matter
(\cite{qui73}, pp. 46-50). For Descartes space is substantial: it is responsable for the
fact that something {\it is} what it is, exactly because of its position in it.
He holds as well that the essential characteristic of material bodies is their {\it
extension}, and that is why in his system matter and space ultimately coincide. In this
way, Descartes remained consistent with the ontological restrictions of the Stagiryte's
logical framework. But the impossibility to show that there can be things with these
primary qualities alone, makes his system untenable\footnote{``A good reason for this is
the fact that we cannot measure primary qualities at all unless we can perceive secondary
qualities. (...) This is the point of Berkeley's argument that a material thing as
conceived by Locke is an impossibility", the Lockean ``primary qualities" of matter being
nothing else than the specification of Descartes' concept of extension. (\cite{qui73}, p.
49.) }. The attempt to develop the possibilities of atomism to a sound philosophical base
for the newly emerging science of nature was undertaken by {\bf Pierre
Gassendi}\footnote{The importance of whom for the development of key-ideas in early modern
science is until today heavily underestimated. It doesn't surprise us much, however, that
Schr\"{o}dinger's finetuned philosophical intuitions recognised this already decades ago.
See (\cite{sch96}, p. 75), also (\cite{jam93}, pp. 34, 92-94); (\cite{blo71}, p. 174).}.
His primal concern was a nominalist critique of the concepts of both Aristotle and
Descartes, a critique which can be accomplished succesfully only by replacing their
logico-ontological categories by ``physical" ones. Aristotle's space was to be rejected
because it was conceived as an {\it accidens}\footnote{In the Peripatetic sense, a quality
that cannot exist apart from the substance to which it belongs. We would say a ``property",
not from a chunk of matter, but from an essential Form.} to substantial form, expressed in
the famous formula that it is ``the number of place". Gassendi was convinced that atomism
provided a tool strong enough to overthrow once and for all the Peripatetic doctrine of
substances and qualities, while avoiding Cartesian ``absurdities" (\cite{blo71}, p. 176)
Hence he radically turned upside down the ontological hierarchy grounded by Aristotle but
was - as Descartes (\cite{qui73}, p. 48) - carefully aware of the necessity to remain
within the ontological constraints of its logical formulation\footnote{Gassendi's work
presents us with one of the rare instances that reveals, both by content and structure,
explicitly the metaphysical nature of the basic categories of modern natural science:
``C'est l'id\'{e}e qui nous para\^{\i}t ressortir de la pr\'{e}sentation de la premi\`{e}re
partie de la Physique dans le {\it Syntagma}, le {\it De Rebus Naturae Universe}. Cette
`Physique' fait imm\'{e}diatement suite \`{a} la `Logique' par quoi commence l'ouvrage
(...). Il ne s'y trouve pas en effet de 'M\'{e}taphysique', et Gassendi s'en explique
d\`{e}s le d\'{e}but: ce n'est pas que la m\'{e}taphysique soit sans objet, ou inaccesible,
{\bf c'est qu'il n'y a pas de distinction entre physique et m\'{e}taphysique} (...). Ce
sont donc bien des cat\'{e}gories physiques qui prennent ici la place de l'ontologie
aristot\'{e}licienne, en m\^{e}me temps qu'elles recoivent un contenu oppos\'{e} \`{a}
celles de la physique d'Aristote. {\bf L'atomisme sera la r\'{e}alisation ad\'{e}quate d'un
tel projet}, mais l'on a vu que celui-ci appara\^{\i}t (...) \`{a} partir de la critique
des `formes substantielles', apporter une nouvelle conception du ``mouvement naturel",
ressusciter 'l'espace des Anciens' contre le `lieu Aristot\'{e}licien',   r\'{e}tablir le
`vide' dans la Nature, proposer une nouvelle notion du ``Temps" etc." (\cite{blo71}, pp.
172-173). Our boldtype.}. That is why he did not content himself with the reformulation of
the ``void" of classical atomism, as it leaves the fundamental question of the origin of
motion unanswered. The introduction of an atomistic explanation for the nature of material
bodies and their properties required a different conception of space as well, for
abolishing the categories ``substance" and ``quality" causes the downfall of the ontology
of ``potentiality" and ``actuality", and therewith of the possibility to grant for change
and motion. Space, infinite, divisible at infinity and indifferent to its material content
(\cite{blo71}, p. 179) is the seat of the stable individuality of things. {\bf Space
constitutes the true Eleactic layer in the metaphysical set-up of natural science:}
things take position in it while it remains immobile and identical to itself. Space
proceeds the material existence of things and is a necessary condition for it, not {\it
vice versa}; it fulfills the same role as the substances with Aristotle. Atomism, Eleatic
by nature, provides a non-qualitative explanation of the composition and motion of material
entities in it. Gassendi's real {\it coup}, however, is the establishment of the rigorous
parallellism between ``absolute space" and ``absolute time". The conclusions arrived at
concerning the ontological status of space will now be applied to time. Time is not the
``number of motion" as with Aristotle. It is motion that depends on time, not {\it vice
versa}. {\bf Time constitutes the true Heraclitean layer in the metaphysical set-up of
natural science}. Gassendi thus reaches the completion of his quest for sound metaphysical
foundations for the new natural science: change {\it is} motion, while space is the {\it
modus of existence} of things permanent, exactly the same way as time is the {\it modus of
existence} of things successive\footnote{``Il faut en effet, comme il le dira plus loin
dans le {\it Syntagma}
\`{a} propos du mouvement, {\bf faire une distinction radicale entre le mode d'existence
des `choses permanentes' et celui des `choses successives', distinction \`{a} laquelle
correspondent respectivement l'espace et le temps}. (...) Espace et temps sont infinis,
l'un selon les dimensions, l'autre selon la succesion (...). Espace et temps ont des
'parties in\'{e}puisables', d'o\`{u} la contingence de la situation du monde {\it hic et
nunc}. Espace et temps sont enfin inalt\'{e}rables et invariables quel qu'en soit le
contenu: ... {\bf l'espace reste identique et immobile, comme le temps s'\'{e}coule
toujours de m\^{e}me mani\`{e}re}." (\cite{blo71}, p.179 sq.), our boldtype.}. Here we find
another fundamental reason for his rejection of the Cartesian substantiality of space:
nobody would defend the substantiality of time. The reality of space {\it and} time is
extra-substantial and extra-accidential. They have dimensions and/or properties, but, being
incorporeal, no functions nor qualities (\cite{blo71}, pp. 177-181). Therefore the
dimensions of space can coincide with the dimensions of matter without causing any
interference. All this clears the road for the theoretical development of the {\it
mathematical} description of the behaviour of material objects in space and time in terms
of their positions and velocities, as had been initiated already by Galilei. The
fundamental paradox however, although neutralised again, did not completely disappear. The
indivisibility of the atom and the infinite divisibility of space somehow break the strict
mutual correspondence between their proper dimensionalities, a necessary condition for
their ontological complementarity. The infinite divisibility of space and
time\footnote{(\cite{moo98}, p. 3). Although we totally agree that ``extension" and
``divisibility" are at the origin of the present-day QM-problems, we think it established
and not without relevance for eventual remedies that these notions are themselves not primitive in this respect, because they follow from past attempts to come to terms with the
paradox.} implies (following Gassendi) continuity. And continuity, that key-concept to
classical mechanics, was the well of which the problems that later led to QM
sprang\footnote{``Our helplessness vis-\`{a}-vis the continuum, reflected in the present
difficulties of quantum theory, is not a late arrival, it stood godmother to the birth of
science"(\cite{sch96}, p.161).}. For {\bf Newton}, Gassendi`s picture provided the perfect
ontology to solve the problems inherent in the mechanical description of nature\footnote{He
uses almost the same wording as Gassendi: ``I. {\bf Tempus Absolutum}, verum, \&
mathematicum, in se \& natura sua, {\bf sine relatione ad externum quodvis, aequalibiter
fluit}, alioque nomine dicitur Duratio (...) II. {\bf Spatium Absolutum}, natura sua {\bf
sine relatione ad externum quodvis, semper manet similare \& immobile} (...)" in
(\cite{new89}, p. 6 (Scholium to the Definitions)). Our boldtype. Reference to this also in
(\cite{pir96}, p. 2). ``Tempus Relativum" and ``Spatium Relativum" are the tools that allow
us to measure and describe motion {\it in concreto}, thence they are easely confused with
`real' - i. e. absolute - space and time.}. However, Newton realised very well that the
force he introduced\footnote{``What he proposed was an addition to the ontology of Nature."
Westfall, R. S., cited in (\cite{coh88}, p. 51). Also (\cite{jam93}, p. 101).} to {\it
explain}  motion needed itself an explanation (\cite{coh88}, p. 40). Already when writing
the {\it Principia} he searched for the explanation of the phenomenon the effects of which
he adequately described therein. For this explanation he turned, as so often, primarily to
the Ancients. Not without disappointment he notes that we do not know of any solution
provided by them (\cite{dob88}, p. 57). After various attempts to develop a sound worldview
on the base of an ``aether" that would allow his ``Force" to be transmitted mechanically
(\cite{dob88}, p. 57-58), he was forced to admit its impossibility, because the effects of
such an aether were not observed on planetary motion. Reluctantly\footnote{In a letter to
Bentley, 25 Feb. 1693, Cited in (\cite{coh88}, footnote 42, p. 52.) } but consequently, he
introduces ``a most subtile Spirit pervading every-body", accounting for both inertial
force as the cause of uniform straight-line motion and gravitation  as {\it actio in
distans}\footnote{In the second edition of the {\it Principia}, which dates 1713:
``Adjicere jam liceret nonulla de {\bf Spiritu quodam} subtillissimo corpora crassa
pervadente, \& in iisdem latente; {\bf cujus vi \& actionibus} particulae corporum ad
minimas distantias se mutuo attrahunt,
\& continguae factae cohaerent; (...) Sed haec paucis exponi non possunt; neque adest
sufficiens copia Experimentorum, quibus leges actionum hujus Spiritus accurate determinari
\& montrari debent. (\cite{new89}, p. 173-174). Our boldtype.}. {\bf Force thus turns out to
be a by necessity immaterial instantiation of space.} Not only is force a physical entity
with a proper ontological status (\cite{del88}, p. 229), but to bridge the abyss between
material and immaterial, between Being and non-Being, this status has to go
beyond the categories of existence of the phenomena it should grant for(\cite{mcg66}, p.
125). It is crucial to realise that Newton {\it needed} the co-incidence of God and space to
account for gravity, in exactly the same way Aristotle needed the First Mover to account for
motion\footnote{``(...) it is possible to see Newton's ideas as
the ``fruition of a long tradition" extending from Aristotle through Newton, a tradition in
which Aristotle's finite plenum was slowly and by painful steps converted into the void,
infinite, three-dimensional framework of the physical world required by classical physics.
Newton's God-filled space was the penultimate development in the process by which concepts
of space were developed by attributing to space properties derived from the Deity; {\bf
after Newton's time, the properties remained with the space while the Deity disappeared
from consideration}." (\cite{dob88}, p. 60.) Our boldtype.}. Post-Newtonian classical
physics however, after being purified by eighteenth century enlightement of all such
``superfluous" hypotheses, disguised by equating force and accelaration the ``causal
paradox" that is the consequence of its metaphysical circularity (\cite{del88}, p. 228).
The persistent problem is the impossibility to {\it prove} the existence of an absolute
``frame of reference" because the immaterial parameters of space and time escape, at least
in principle, experimentally controllable observation. It is exactly this ontological
status {\it beyond} that makes ``similar and immobile" space to the only warrant for the
repeatability of experimental observations (\cite{jam93}, p. 84), which is another way to
highlight its Eleactic character, and its enigma as well. The essentially Leibnitzian
approach of what was called later ``analytical mechanics" seemed to deliver an acceptable -
because mechanical - alternative, while allowing at the same time for the treatment of
similar classes of physical problems as ``Newtonian" or ``vectorial" mechanics
(\cite{lan70}, p. xxi). The {\it principle of least action} presents a description of
mechanical systems by minimizing a quantity that measures the ``action" of the system under
consideration {\it as if} it were a single particle moving in a {\bf plenum}, the particles
of this plenum remaining mutually seperated, but being always in direct contact with each
other. The mathematical ``configuration space`" represents space {\it as if} force were the `real'
manifestation of its n-dimensional geometry (\cite{lan70}, p. 13). But this at first glance
purely formal approach hides the Cartesian-Leibnizian ontology of the unreality of space,
with Descartes because space coincides completely with matter as pure extension; with
Leibniz because space is nothing but the relations between the objects `in' it. That the
basic quantity describing with {\bf Leibniz} the dynamical behaviour of the system, the
{\it vis viva} or living force\footnote{Almost identical with our kinetic energy. Together
with the ``work of the force" or potential energy the two fundamental scalar quantities on
which the study of equilibrium and motion of analytical mechanics rests (\cite{lan70}, p.
xxi).}, is represented by a {\it scalar} is clearly of more than merely formal importance,
it is a consequence of his ontological position. This annihilation of non-Being - the {\bf
vacuum}, empty space - from physics opens the possibility to treat space as a merely
``relativistic" phenomenon, a position fully supported later by Einstein in the context of
his Special Relativity\footnote{``The introduction of a ``luminiferous ether" will prove to
be superfluous inasmuch as the view here to be developed will not require an ``absolutely
stationary space" provided with special properties, nor assign a velocity-vector to a point
of the empty space in which electromagnetic processes take place." (\cite{ein52}, p. 38).
}. But the problem of the initial {\it origin} of forces or motions remains equally
unsolved. Even worse, analytical mechanics lacks a `natural' way to provide for the stabilising
frame of reference for experimental observation as is present in the explicit ontology of
the Newtonian treatment. Appearances are saved by stating that the universe presents the
same aspect from every point (apart from ``local irregularities") (\cite{jam93}, p. 84), but
in the ontological setting of the logical framework - i.e., the framework of the separation
between Being and non-Being, between system and environment, between cause and effect -
this can be uphold only by an act of Divine creation, as is the case with Leibniz, but of
course not with his followers in the later school of analytical mechanics. Saying that the
notion of absolute space is redundant {\it because} the natural laws are invariant under
coordinate transformations (\cite{har95}, p. 1) is turning upside down the chain of
justifications. As we hope to have made clear by now, the whole metaphysical set-up of
natural science is such that they {\it should} be so. The fact that the role of ``causal
black box" played by space is now fulfilled by the ``environment"\footnote{``What is
system, for instance, is described by phases or states; environment is not, and cannot, be
represented in such terms. Rather, {\bf environment is the seat of ``external"} {\it
forces} (...). This apparently necessary and innocent partition of the world into system
and environment, (...) has the most profound consequences for the notion of causality. For
according to it, the notion of causality becomes bound irrevocably to what happens in
system alone, (...) is the state-transition sequence. (...) {\bf what happens in
environment has thus been put beyond the reach of causality. Environment has become
acausal}." (\cite{ros89}, p. 19). Our Boldtype.}, doesn't change anything to the
fundamental circularity of the reasoning.

\section{The Re-emergence of the Paradox: Late Modernity} 
 
The dept of the problem has been brought again to the fore by the results obtained both
theoretically and experimentally by QM. For not only does the theory incorporate states for
quantum entities that imply non-local behaviour \cite{aerent}, but such effects have by now
been established incontrovertibly in various experimental settings, thus shortcutting for
once and for good eventual attempts to explain them away
statistically\footnote{C. Piron in \cite{pir85}: ``... the orthodox QM (the `new
testament' as Pauli named it) with its credo and its principles is dead and
definitively dead." A lot of acrobatic attempts where made to rescue the
modernist picture, by extending the quantum picture with an underlying modernist
kind of world (for example D. Bohm
\cite{boh52}), or by rejecting it (for example A. Einstein \cite{ein35} and J.
Bell 
\cite{bel65}). But at the same time, every advance on an experimental level (in
particular A. Aspect
\cite{asp82} and H. Rauch \cite{rau}) confirmed that these attempts were bound to
fail. The general community of physicists reacted to this by a kind of {\it
ontological ignorance} attitude which evolved towards a {\it pragmatic quasi
botanic empiricism}, or by some {\it new age like mystification tendencies} (see
for example `The Tao of Physics' by F. Capra). Alternatively, a not unimportant group
of people saw the inadequate arsenal of mathematical (and in particular logical)
tools as the origin of all these problems (especially J. Von Neumann $\&$ G.
Birkhoff \cite{bir36}, J Jauch
\cite{jau68}, C. Piron \cite{pir76}, H. Neumann
$\&$ G. Ludwig \cite{lud81a,lud81b} and D. Foulis
$\&$ C. Randall \cite{fou78}).  This gave rise to an enormous development of mathematical
attributes and structures, of which {\it quantum logic} is the most known among philosophers
of science (see for example \cite{bel81} and in particular \cite{mit81}).  Unfortunately,
the mathematical expertise required made the development
and study of quantum logic an essentially mathematical occupation such that the conceptual
development stagnated (a confirmation of this fact can be found in
\cite{moo93} and \cite{pir90}).}. Furthermore it was proved to be impossible to describe
soundly two or more separated systems within the present formalism \cite{aer81}.
As is sufficiently known, the state of a physical system in QM is described
within mathematical Hilbert space. For every physical entity the collection of all its
properties constitutes its state in Hilbert space. Exactly in the same way
as the state of a classical entity is represented in phase space, the Hilbert space contains all
possible states the system can possibly have. In this sense, both phase space and the
Hilbert space represent the {\it environment} in which the dynamical transition of one state to
another of the entity will take place. To avoid that the individual state be blurred in the
statistical ensemble proper to the `orthodox' interpretation, C. Piron introduces a new notion of
physical state based on the concept of property as related to experimental projects with
well-defined certain results (\cite{pir81}, p. 398). These results are called yes-results, while
any other outcome is considered to be a no-result. To this end, a test that could
{\it eventually} be performed is associated to a property of the system under
consideration, in such a way that, once you are certain {\it in advance} to
obtain the desired result, you can assign to the system an {\bf element of
reality} (\cite{pir76a}, p. 170) as conceived by Einstein (\cite{ein83}, p. 137).
Introducing Aristotle's dynamical terminology, a property is called ``actual"
when the result of the test is certain; when uncertain it is said to be
``potential" (\cite{pir76a}, p. 171). A property conceived this way is always
potential or actual. It is however possible that the property tested by an
experimental project, and the one tested by the experimental project obtained by
exchanging the ``yes" and the ``no" results of the former, are {\it both} potential.
Whenever this situation appears we are dealing with quantum-like entities,
contrary to classical ones where this can never be the case. But it is not {\it a
priori} obvious how to define a particular entity and thus how to assign
properties to it. All comes down to account for the set of tests which matches
the collection of its properties and thus the entity itself. This presupposes
the possibility to separate the phenomenon under consideration from the rest of
the universe \cite{dan89}. But, although it is stated explicitly that the
certainty of the experimental project is an objective feature of the system
without reference to our knowledge or beliefs (\cite{moo98}, p. 5), we are
supposed to know in advance what to do with the system to get this certain
result. Indeed, the system should be prepared in a precise way that is related
to our existing knowledge of its properties. This presents no problem when we
are dealing with phenomena that are accessible to our daily experience, like
with the breaking of a piece of chalk (\cite{pir85}, p. 208.). The necessary
assumptions about its properties are then given by our daily experience. This enables
us to drop the preparation procedure that is essential to `true' experimental
observation. Experimental observation is indeed more than the ``contemplation
of the astronomer" (\cite{moo98}, p. 8), it is an intervention that {\it
prepares} the system in such a way that we {\it obtain} an entity - separated
from its environment - together with a set of properties determined by the
possible yes-no experiments. Again, it is {\it the preparation of the system} that
moulds the `real thing' so as to fit in a definite metaphysical set-up,
primarily by separating ``the system" out of its ``environment". Whatever be the
structure of the set of possible answers, the modelisation of the set of
possible yes-no experiments imposes on the set of possible properties the
mathematical structure of a complete lattice (\cite{jau69}, p. 844). This means,
as indicated before, that {\bf the ``real thing" is forced to fit into the
scheme of an intrinsically Eleatic ontology. The preparation procedure is a
`black box' generating that ontology}. This does not imply that it is senseless
to perform experimental projects. Natural laws are verified by experimental
observation, and experiments are performed in reality, so there must be some
kind of agreement with what happens in reality. But it is precisely the {\it
kind} of relationship that exists between reality and experimental observation
that should be rendered more clear; which implies that the consequences of the
fact that each experimental project is an {\it intervention} in reality should
be itself subject of investigation. By making explicit the structures underlying
the sets of possible yes-no questions and of possible answers, the Geneva School
approach realises an important progress in this respect, though until now the
ontological aspect of the problem has been left untouched. Embarking on an
analysis of this aspect could nevertheless shed some light not only on the
relation between the ``entity" and the ``real-world phenomenon", but also on the
status of our conceptions of space and time, which form, as has been argued, the
ontological foundation for the metaphysical set-up present in scientific theory.
This provides the background against which Piron's attempts to get hold of a
``realistic" conception of space are to be seen. The inevitable
re-appearance of ontological circularity to which also these attempts are
subject can be clarified on a more theoretical level\footnote{A specific
discussion of this problem with respect to Piron's approach of space and time
will be the subject of a future publication. Equally problematic but outside our
scope would be an analysis of the concept of `field', arising out of General
Relativity. Be it sufficient to refer to Schršdinger, who explains: ``at any
rate the very foundation of the theory, viz. {\bf the basic principle of
equivalence of acceleration and a gravitational field}, clearly means that there
is {\bf no room for any kind of `force' to produce accelaration save
gravitation, which however is not to be regarded as a force but resides on the
geometry of space-time}." (\cite{sch85}, p. 1). This `making real' of the
geometry of space as the `force' behind dynamical processes in four-dimensional
space-time reveals more clearly the underlying ontology and links General
Relativity firmly to the `Leibnizian' tradition within physics. But this
continuum, in which gravity manifests itself, is itself the source of a host of
ontological problems. Regarding this, Piron cites in his lecture Einstein saying
``that according to general relativity {\bf space is endowed with physical
qualities and in this sense an ether exists} (...) but this ether must {\bf not
be thought of as endowed with the properties of ponderable media (...) nor may
any concept of motion be applied to it}." (\cite{pir97}, p. 1, 2). Our boldtype.
By trying to explain his own concept, Einstein reveals how close Newton and
Leibniz actually stand.}. The tacit assumption underlying the position that a
thing coincides with a collection of properties is that ``being" in the sense of
``existence" is but a property amongst the others. We discussed before at length
why this assumption is untenable in the context of an - implicit or explicit -
Eleatic ontology as the one present in the cognitive and empirical procedures of
science. The notions ``object" or ``entity" {\it necessitate} the introduction
of Being in some form, ontologically {\it beyond} the object's predicable
properties, so as to grant for its required stability and separability, as well
as for an origin for the dynamics governing the processes of change and motion
to which it is exposed. In this ontological setting {\bf to be is not a
property}. It follows that existence is not a predicate (\cite{qui73}, p. 36)
and existence itself is not within reach of experimental preparation, nor
observation (\cite{jam93}, p. 84). Regarding the paradoxical results attained by
QM, we are led then to the conclusion that only the development of an explicit
and properly ontological foundation for basic categories as `entity', `motion'
as well as for the intervention represented by their experimental observation
remains open as a fruitful strategy. A sketch of a possible further elaboration
of the results attained thus far by the Geneva School with regard to
the foregoing analysis is presented in the last section of this paper. 
Alternatively, leaving behind the principle of contradiction as constituting
ontology at the core of scientific procedure both empirical and cognitive, could
be taken into consideration. Interesting steps to come to terms with problems
related to this approach have been made in the recent past. From attempts to
develop logical systems that escape the rigorous consistency demanded by the
principle of contradiction sprang relevant considerations, but these remain
trapped within the framework of the ontological core common to {\it all} logical
systems\footnote{``Every logician in the end divides propositions into those
which are acceptable and those which are not." (\cite{eps90}, introduction)}.
Another approach recognises the {\it ontological} nature of the problem, but
limits itself to the construction of ontologies for ``possible worlds" that fit
``a" logic in which the principle of contradiction has been ``neutralised"
somehow \cite{res80}. Since all logics still do partake in the same fundamental principle,
these ``possible worlds" necessarily remain Eleatic. Our position would be that the origin
of the problem resides {\it within the ontological nature of reality itself}. This viewpoint might
clarify the incurable presence of paradoxes in logical systems of all kinds. Paradoxes then
would appear because of an ontological incompatibility between logic and ``real" reality.
Therefore our approach would be to take {\it paradox itself} as the starting point for the
construction of an ontological framework. We see the work of Spencer Brown \cite{spe69},
and, more recently, Kaufman\footnote{Oral communication; see also \cite{kau95}.} as
supportive in this respect. That this strategy might open relevant perspectives for QM is
indicated by the recent discovery of Aerts et al
\cite{aer98b}. The principle underlying such ontology has been adequately formulated
in the past. It is the already mentioned {\it principle of coincidence of opposites}, to be
found in the work of fifteenth century philosopher Nicolaus Cusanus. It provides an excellent 
example to show that abandoning the
principle of contradiction implies the loss of neither the capacity to reason soundly, nor
the possibility to use mathematics (\cite{cus32}, viz.
Capitula XII to XVII). It does imply, however, the necessity to abandon on a
conscious level the artificial separation between ``subject" and ``object", the distance
between things being a mere aspect of their instantiation. As an indication for its
operational viability we suspect the possibility to consider circularity and
self-reflexivity as tools rather than as problems. Another interesting possibility might be
the development of a sound conception of ``physical space", homogeneous from all
points, because circumference and center coincide in all points, and without the need to
introduce a force-like ``First Mover", since distance and interaction reveal themselves as
the same thing. This again might add to opening up interesting perspectives on the
``non-locality" of elementary physical phenomena manifest at the core of QM. 
 
\section{Contemporary physics as products of early Greek thought: Post-Modernity and beyond?}

Let us now concentrate on one of the words in the title which might seem rather innocent and
obvious, namely `products'. In a first reading of this title one could think that it
might easily be replaced by `result' or `onsequence'. We nevertheless attach a very definite
meaning to this word 'product' which is definitely not covered by `result' nor `consequence'.  In
contemporary mathematics, which deals with the study of families of well-defined formal structures
(or, in a more advanced language, `categories of mathematical  objects'\footnote{For an outline of
the mathematical theory of categories introduced by
Eilenberg and Mac Lane \cite{eil45} we refer to
\cite{bar95,mac72,mcl95}. In fact, some approaches and authors on category theory see a
possibility of introducing `aspects of undefinedness' within this formalism, and consider
this aspect as the main argument to apply category theory as a foundation for mathematics.}), 
products of two structures that belong to the same family (i.e., two mathematical objects
that belong to the same category) are usually defined as a third structure in this same
family which has the two given structures as `faithful' substructures (in categorical
language this means that there exist structure preserving maps, called `morphisms', from the
two given objects into the product object). In fact, this means that the product is larger
than the given structures, but essentially not different. This way of defining products is
an obvious consequence of the specific `scientific method' in the context of which
mathematics is used: one only wants to deal with objects that are well defined. Once one has defined
such a well-defined collection of objects, one sticks with it and studies it. Even more, an object
has only a meaning as an element of this well-defined collection. As a consequence, if one
wants to introduce operations on the objects in this collection in order to study and
characterize it, one remains faithful to this collection, i.e., one does not leave
it\footnote{We have to remark that quite recently, since the development of the mathematical
theory of categories, one became very interested in `relating' the different well-defined
collections of mathematical objects. However, one is still attached to the same kind of
scientific method, but now on a meta-level.}. This situation in mathematics had a
major influence on the way how people tend to describe compound systems in physics: for a
given description (i.e., structural characterization) of individual physical entities one is
still attached to the idea that the compound system should be described within the same
family of structural characterizations. Within the context of Newtonian mechanics, this
does not pose a problem. Within the context of QM however, one is able to produce a description
procedure for the joint system, but looses completely the mathematical
consistency\footnote{This inconsistency is encountered as well on a purely structural
level, in the sense that there exist two unequivallent procedures for obtaining a so called
tensor product (see \cite{aerdau}), both when one starts from an operational or
empiricist point of view (see \cite{aer81}, \cite{pir90} and \cite{fra}).} required for a
`good theory'. Unfortunately, by yielding the conclusion that the `Hilbert space
structure', the structure in which QM-entities are described, is not `the good
universal structure' for their description, these results have only fostered the
search  for `bigger' and `bigger' structures in which one could hope to find `consistent products'
for the description of the compound system. Never the {\it a priori} idea of universality
has been put into question. 
According to our previous sections, we think that this specific attachment to universality
of a description points the finger at the real problem at the source of this malaise
concerning the description of compound entities in physics.  {\it The presence of a second entity
in a compound system definitely changes the context of the first one}, and as such, a
description that incorporates a contextual ingredient should take this presence into account
in an explicit way\footnote{For the reasons stated before, this could considerably
enhance the coherence of the theory.}. Unfortunately, for the QM formalism, this is not
the case\footnote{An explicit construction in which one of the present authors tries to take
the context into account for a situation of `the many' (see
\cite{HxHx...,PIcat}), based on ideas developed by Aerts, Gisin and Piron in
\cite{aer86,aer94} and \cite{gispir}, and yielding the framework of \cite{aer94} and \cite{coe97},
shows that in the traditional Q.M. description, aspects like the order in which one performs
consecutive measurements on different individual pseudo-entities within the compound system are
even not a part of the context, but an ingredient of the formal representation of the entity
itself.}. Solving this problem comes down to starting from correct collections of primitive
notions in order to reconstruct the proper formalization for the concept under consideration, in
this case the concept of compoundness\footnote{A more detailed discussion on this matter can be
found in
\cite{moo98}.}. With the goal to aim to an as reduced as possible reformulation of the
formalism that arises when one starts from as few as possible primitive notions, Moore
started {\bf a categorization program\footnote{This program, when consistently developed in the
total framework of the theory, could amount in an explicit Aristotelian ontological set-up, exactly
in the sense Piron \cite{pir76a} intended.} for the foundations
of physics} (see
\cite{moo95,moo97,moo98}). Within this scheme, other primitive notions have been incorporated (see
\cite{amcoe,bofr,coemoo,Boston,PIcat}), yielding a categorical description of the concept of
compoundness (see \cite{PIcat}). Without falling into the trap of empirical
fragmentation\footnote{Which occurs as wel on a theoretical as on a operational level.}, it seems
to be possible for this concept to skip the universality principle which has always been the a
priori for theoretical physics. One could wonder whether other physical concepts could be treated
in the same way?  A conclusive question?  When dealing with 'questions' we are confronted again
with the problem of definedness. Old and endless debates on the yes or no necessity 
of explicit representations of 'states' of reality could definitely be omitted by allowing
aspects of undefinedness within the explicitation of a concept such as states and experimental
projects; quoting Moore on the later in \cite{moo98} p.3:
{\it "I shall take a particular physical system to be some part of the ostensively external
phenomenal world, supposed separated from its surroundings in the sense that its interactions with
the environment can either be ignored or effectively modeled in a simple way. The restriction to a
reasonably circumscribed aspect of a phenomenon is crucial: physics is, first and foremost, the
study of detailed models of specific situations ...
I shall take a definite experimental project relative to a physical system to be a real
experimental procedure where we have defined in advance what would be the positive response should
we perform the experiment ... 
Of course the specification of a precisely delimited positive
response is also an idealization: in general terms, all that we require from an assignment rule is
that it be sufficiently clear in each case whether or not one should assign the response 'yes',
this within subjectively reasonable limits with respect to borderline cases"};
where he motivates the necessity of consecutive and complementary 'idealizations' by: 
{\it "... the problem of linguistic demarcation in the face of vagueness or contextuality."}
As it is very reasonable to consider human creation as
an ability to 'invent' new objects outside an {\it a priori} given class $\{x|\phi(x)\}$ with a
characterizing predicate form
$\phi$, the collection of possible designable experimental projects cannot be a class and in
particular, cannot be a set.  To quote Moore (private communication, 27-09-97):
{\it "... this boils down to a kind of object-subject dichotomy: states
and properties are about the exterior world and definite experimental projects
are about the internal world. Building physical theories is a fallible
attempt to relate one to the other ..."}   
Still, any operational theory attempts to remain unaware, by formal ignorance, of
these possible creative acts consisting of designing definite experimental projects. This
ignorance could be motivated by the ever ruling set theoretical basis for mathematics.
In order to understand  the problems that arise, one could consider entities like {\it my past} or
{\it the entity's past}, also studied in \cite{Boston}. It is obviously impossible to define
a predicate
$\phi$ that characterizes all my possible pasts that I can have on a next instance of time, since
this would require Laplace's supreme intelligence: having all knowledge on possible events that
might be imposed by the context between now and the considered next instance of time.  Contrary to
the existence of a set of states as a criterion for the reality of an entity, the existence of
properties as reference to the elements of reality in the Geneva approach  
\cite{aer81, moo95, moo98} can be the starting point for a more flexible setup: the entity 'my
past' for example can be {\it described by consecutive creations of properties, themselves not being
part of a predefined set}.  
This idea has been generalized and taken as a starting point of of a formal scheme refered to as
the {\bf Induction\footnote{Here, 'induction' does not refer to any 
of its usual philosophical significances, but rather refers to physical theories like
electro-magnetism.} formalism}, which takes in an {\it a priori} (and not only in an explanatory)
way 'creation' into account, as well 'human' creation (choice/invention) as 'material' creation
(mutual Induction as interaction), and of which a complete technical development can be found in
\cite{amcoe,Boston,QmInd,PIcat}. This approach succeeds in avoiding the aspects of isolation in
the definition of an entity: the 'particular physical system' fuzzyness to which Moore referes is
explicitely not present. As such, this strategy aims at the elaboration of the results of the
Geneva school in the direction of a formalism that is explicitly compatible with the results
of the foregoing ontological analysis. 
Formally as well as conceptually, such a theory of creation, by
constituting the description of the emerging properties of the entities as possible parts
of compound systems, treats
``interaction with the context of an entity as a part of a measurement process  (cfr.
\cite{aer86,aer94,coe97,bofr})" and ``entanglement of individual entities  (cfr.
\cite{HxHx...,PIcat})" on the same level, i.e., in terms of mutual Induction of
properties (cfr.
\cite{amcoe,Boston,PIcat}). We finally want to remark that the idea of going beyond definedness,
going beyond set-large collections of states, seems to be most conveniently expressed through the
use of so-called quasi-categories (cfr. \cite{ada90} p.31), a generalization of the above mentioned
mathematical categories, of which the undeniable necessity thus emerges from both physical and
metaphysical considerations. To quote Hersh in
\cite{her86} p.22-23:    
{\it "... We can try to describe mathematics, not as our inherited prejudices imagine it to be,
but as our actual experience tells us it is. ... What are the main properties of mathematical
knowledge, as known to all of us from daily experience? (1) Mathematical objects are invented or
created by humans. (2) They are created, not arbitrarily, but arise from activity with already
existing mathematical objects, and from needs of science and daily life  ..."}   
 
\section{Acknowledgments.}    

We thank Sonja Smets, Hubert Dethier, Wilfried Van Rengen and Rudolf
De Smet for discussing the content of this paper.  We thank David Moore for the indication of
problems and references connected to the subject of this paper.  We thank D. Durlinger for
technical help while implementing the ancient Greek font.  Bob Coecke is Post-Doctoral Researcher
at Flanders' Fund for Scientific Research.


\begin{thebibliography}{99}
 
\bibitem{ada90}
J. Ad\'amek, H. Herrlich and G.E. Strecker, {\it Abstract and Concrete Categories}, J. Wiley \&
Sons (1990).

\bibitem{aerdau}  
D. Aerts and I. Daubechies, {\it Helv. Phys. Acta} {\bf 51}, 661 (1978).
 
\bibitem{aer81}  
D. Aerts, {\it The One and the Many}, PhD-thesis, Free University of Brussels (1981); {\it
Found. Phys.} {\bf 12}, 1131 (1982).

\bibitem{aer86}  
D. Aerts, {\it J. Math. Phys.} {\bf 27}, 202 (1986).

\bibitem{aer94}  
D. Aerts, {\it Found. Phys.} {\bf 24}, 1227 (1994).
  
\bibitem{aer98a}  
D. Aerts, B. Coecke and S. Smets, 'On the Origin of Probabilities in Quantum Mechanics:
Creative and Contextual Aspects', in G. Cornelis and J.P. Van Bendeghem, eds., {\it Einstein
Meets Magritte: Metadebates}, Kluwer Academic Publishers (1998).
 
\bibitem {aer98b} 
D. Aerts, J. Broeckaert, S. Smets, 'A Quantum Mechanical Description of
the Liar Paradox', submitted for the
proceedings of the 20th Int. conf. on Philosophy, Boston (1998).

\bibitem{aerent}  
D. Aerts, 'The Entity in Modern Physics: the Creation-Discovery-View of Reality', in E.
Peruzzi, ed., {\it Identity and Individuality of Physical Objects}, Springer Verlag, Berlin
(1998).

\bibitem{amcoe}
H. Amira, B. Coecke and I. Stubbe, 'How Quantales Emerge by Introducing Indiction in Operational
Theories', Preprint, FUND-DWIS, Free University of Brussels (1998).

\bibitem{are78} 
H. Arendt, {\it The life of the mind. One / Thinking}, one-volume edition,
Harcourt Brace Jovanovich Publishers, London (1978).

\bibitem{asp82}
A. Aspect et. al., {\it Phys. Rev. Lett.} {\bf 49}, 1804 (1982).

\bibitem{bar95}
M. Barr and C. Wells, {\it Toposes, Triples and Theories}, Springer (1995).

\bibitem{bel65}
J. Bell, {\it Physics} {\bf 1}, 195 (1965).

\bibitem{bel81}
E.G. Beltrametti and B.C. van Fraassen, eds., {\it Current Issues in Quantum Logic}, 
Plenum (1981).

\bibitem{bir36}
G. Birkhoff and J. von Neumann, {\it Ann. Math.} {\bf 37}, 823 (1936).

\bibitem{blo71} 
O.R. Bloch, {\it La philosophie de Gassendi. Nominalisme, Mat\'{e}rialisme et
M\'{e}taphysique}, Martinus Nijhoff, La Haye (1971).

\bibitem{boh52}
D. Bohm, {\it Phys. Rev.} {\bf 85}, 166 (1952).

\bibitem{cam91} 
J. Campbell, {\it The Masks of God, volume I},  Primitive Mythology, Arkana,
Penguin Books, N.Y. (1991).

\bibitem{car92} 
R. Carnap, {\it Empiricism, Semantics, and Ontology}, in R. Boyd, P. Gasper and J.D. Trout, eds.,
{\it The Philosophy of Science}, MIT press, Massachusetts (1992).

\bibitem{cas73} 
H.N. Castaneda,  {\it Plato's theory of relations}, in M.
Bunge, ed., {\it Exact Philosophy}, D. Reidel Publishing Company, Dordrecht (1973).

\bibitem{coe97}  
 B. Coecke, {\it Hidden Measurement Systems}, Doctoral Dissertation,
Free University of Brussels (1995); {\it Helv. Phys. Acta} {\bf 70}, 442 (1997);
{\it Helv. Phys. Acta} {\bf 70}, 462 (1997).

\bibitem{HxHx...}   
B. Coecke, 'A Representation for Compound 
Quantum Systems as Individual Entities: 
Hard Acts of Creation and Hidden Correlations' and 'A Representation for Spin-S Quantum
Entities as 2S Individual Spin-1/2 Quantum Entities', {\it Found. Phys.}, to appear (1998).

\bibitem{bofr}  
B. Coecke and F. Valckenborgh, {\it Int. J. Theor. Phys.} {\bf 37}, 311 (1998).

\bibitem{coemoo} 
B. Coecke and D.J. Moore, 
'Decompositions of Probability Measures on Complete Ortholattices in Join Preserving 
Maps', Preprint, FUND-DWIS, Free University of Brussels (1998).

\bibitem{Boston} 
B. Coecke and I. Stubbe, 'On a Creation-Based Extension of Operational Theories', submitted for the
proceedings of the 20th Int. conf. on Philosophy, Boston (1998).

\bibitem{QmInd} 
B. Coecke and I. Stubbe, 'Quantum Mechanics as a Theory of Induction',
Preprint, FUND-DWIS, Free University of Brussels (1998).

\bibitem{PIcat} 
B. Coecke, 'Compoundness as Diagrams in Quasi-Categories',
notes of a lecture given at IQSA-Atlanta, August 1997; B. Coecke and I. Stubbe, 'The Category of
Mutual Inductions for Entities Described by Pseudo States', In preparation (1998).

\bibitem{coh88} 
B.I. Cohen, 'Newton's third law and universal gravity', in P.B. Scheurer and G. Debrock, eds., {\it
Newton's scientific and philosophical legacy}, Kluwer academic
publishers, Dordrecht (1988).

\bibitem{con87} 
M. Conch\'{e}, 'Epicure. Lettres et maximes', texte \'{e}tabli par Marcel
Conch\'{e}, PUF, Paris (1987).

\bibitem{cus32} 
De Cusa, N., {\it De Docta Ignorantia}, in E. Hoffmann and R. Klibansky, eds.,
Lipsiae, In Aedibus Felicis Meiner (1932).

\bibitem{dan89} 
W. Daniel, {\it Bohr, Einstein and Realism}, Dialectica 43 (1989).

\bibitem{del88} 
E. Dellian, 'Inertia, the innate force of matter', in P.B. Scheurer and G. Debrock, eds., {\it
Newton's scientific and philosophical legacy}, Kluwer academic publishers,
Dordrecht (1988).

\bibitem{die71} 
H. Diels and W. Kranz, {\it Fragmente der Vorsokratiker, erster Band}, Weidmann,
Dublin, Z\"{u}rich, 1971.

\bibitem{dij96} 
E.J. Dijksterhuis, {\it De mechanisering van het wereldbeeld}, 7th ed.,
Meulenhof, Amsterdam (1950-1996).

\bibitem{dob88} 
B.J.T. Dobbs, 'Newton's alchemy and his `active principle' of gravitation',
in P.B. Scheurer and G. Debrock, eds., {\it Newton's scientific and philosophical legacy}, Kluwer
academic publishers, Dordrecht (1988).

\bibitem{ein35}
A. Einstein, E. Podolski and N. Rosen, {\it Phys. Rev.} {\bf 47}, 777 (1935).

\bibitem{ein52} 
A. Einstein, 'On the electrodynamics of moving bodies', in {\it The principle of
relativity. A collection of original papers on the special and general  theory of
relativity. Notes by A. Sommerfeld.}, Dover publicatons, New York (1952).

\bibitem{ein83} 
A. Einstein, 'Can quantum-mechanical description of physical reality
be considered complete?', in J.A. Wheeler and
H.Z. Wojciech, eds., {\it Quantum Theory and Measurement}, Princeton University Press, Princeton (1983).

\bibitem{eil45}
S. Eilenberg and S. Mac Lane, {\it Trans. Am. Math. Soc.} {\bf 58}, 231 (1945).

\bibitem{eps90} 
R.L. Epstein, {\it The semantic foundations of Logic}, Vol. I, Kluwer academic
publishers, Dordrecht (1990).

\bibitem{esp76}
B. d'Espagnat, {\it Conceptual Foundations of Quantum Mechanics}, W.A. Benjamin (1976).

\bibitem{fou78}
D.J. Foulis and C.H. Randall, 'Manuals, Morphisms and Quantum Mechanics', in H. Marlow,
ed., {\it Mathematical Foundations of Quantum Theory}, Accademic Press. (1978).

\bibitem{foup83}
D.J. Foulis, C. Piron and C.H. Randall, {\it Found. Phys.} {\bf 13}, 813 (1983).

\bibitem{fou83}
D.J. Foulis and C.H. Randall, {\em Found. phys.} {\bf 13}, 843 (1983).

\bibitem{fow67} 
H.N. Fowler, {\it Plato in twelve volumes}, Loeb Classical Library, Heinemann,
London (1967).

\bibitem{fri93} 
K. Von Fritz, 'Nous, noein, and their derivatives in pre-Socratic
philosophy', in A.P.D. Mourelatos, ed., {\it The Pre-Socratics, a colletion of critical essays},
Princeton University Press, Princeton (1974-1993).

\bibitem{gispir}
N. Gisin and C. Piron, {\it Lett. Math. Phys.} {\bf 5}, 379 (1981).

\bibitem{har95} 
R. Harr\'{e}, 'The redundancy of Spacetime: Relativity from Cusa to
Einstein', in D. Aerts, ed., {\it Einstein Meets Magritte: White Book}, Kluwer Academic
Publishers (1998).

\bibitem{her86}
R. Hersh, in T. Tymoczko, ed., {\it New Directions in the Philosophy of Mathematics} p.9,
Birkh\"auser (1986).

\bibitem{jam93} 
M. Jammer, {\it Concepts of Space. The history of theories of space in physics},
3th ed., Dover Publications, N.Y. (1954-1993).

\bibitem{jau63}
J.M. Jauch and C. Piron, {\it Helv. Phys. Acta} {\bf 36}, 827 (1963).

\bibitem{jau68}
J.M. Jauch, {\it Foundations of Quantum Mechanics}, Addison-Wesley (1968).

\bibitem{jau69} 
J.M. Jauch and C. Piron, 'On the structure of quantal proposition systems',
Helv. Phys. Acta {bf 42} (1969).

\bibitem{jay76} 
J. Jaynes, {\it The origin of consciousness in the breakdown of the bicameral
mind}, Houghton Mifflin Company, Boston (1976).

\bibitem {kau95} 
L. Kaufman, 'Virtual Logic', in D. Aerts, ed., {\it
Einstein Meets Magritte}, Kluwer Academic Publishers (1998).

\bibitem{kir75} 
G.S. Kirk, {\it Heraclitus. The Cosmic Fragments}, Cambridge University Press,
Cambridge (1975).

\bibitem{kir83} 
G.S. Kirk, J.E. Raven and M. Schofield, {\it The Presocratic Philosophers. A
critical history with a selection of texts}, Cambridge University Press, Cambridge (1983).

\bibitem{lan70} 
C. Lanczos, {\it The variational principles of mechanics}, fourth ed., Dover publications,
New York (1970).

\bibitem{luc85} 
G. Luck, {\it Arcana Mundi. Magic and the Occult in the Greek and Roman Worlds. A
collection of Ancient Texts}, The Johns Hopkins University Press, Baltimore and London (1985).

\bibitem{lud81a}
G. Ludwig, 'An Axiomatic Basis of Quantum Mechanics', in H. Neumann, ed., {\it
Interpretations and Foundations of Quantum theory}, B.I.-Wissenschaftsverlag (1981).

\bibitem{lud81b}
G. Ludwig and H. Neumann, 'Connections between Different Approaches to the Foundations of
Quantum Mechanics', in H. Neumann, ed., {\it Interpretations and Foundations of Quantum
theory}, B.I.-Wissenschaftsverlag (1981).

\bibitem{mac72}
S. Mac Lane, {\it Categories for the working Mathematician}, Springer (1971).

\bibitem{mcg66} 
J.E. McGuire and P.M. Rattansi, 'Newton and the `Pipes of Pan'', {\it Notes
and records of the Royal Society of London} {\bf 21} (1966).

\bibitem{mcl95}
C. McLarty, {\it Elementary Categories, Elementary Toposes}, Oxford Sci. Publ. (1995).

\bibitem{mit81}
P. Mittelstaedt, 'Classification of Different Areas of Work Afferent to Quantum Logic', in
E.G. Beltrametti and B.C. van Fraassen, eds., {\it Current Issues in Quantum Logic}, Plenum
(1981).

\bibitem{moo93}
D. Moore, {\it Helv. Phys. Acta} {\bf 66}, 471 (1993).

\bibitem{moo95}
D. Moore, {\it Helv. Phys. Acta} {\bf 68}, 658 (1995).

\bibitem{moo97}
D. Moore, 'Closure Categories', {\it Int. J. Theor. Phys.}, to appear (1997).

\bibitem{moo98}
D. Moore, 'On State Spaces and Property Lattices', {\it British J. Phil. Sc.}, to appear
(1998).
 
\bibitem{mou93} 
A.P.D. Mourelatos, {\it The Pre-Socratics, a collection of critical essays},
Princeton University Press, Princeton (1974-1993).

\bibitem{new89} 
Newton, I., {\it Isaaci Newtoni, Opera quae extant omnia}, Commentariis illustrabat Samuel
Horsley, Londini: exc. Joannes Nichols (1889).

\bibitem{owe71} 
G.E.L. Owen, 'Plato on Non-Being', in G. Vlastos, ed., {\it Plato: a
collection of Critical Essays}, vol I, Doubleday (1971).

\bibitem{pir76}
C. Piron, {\it Foundations of Quantum Physics}, W.A. Benjamin (1976).

\bibitem{pir76a} 
C. Piron, 'Le r\'{e}alisme en physique quantique: une approche selon
Aristote', in {\it The concept of physical reality. Proceedings of a conference organized by the
Interdisciplinary Research Group}, University of Athens (1983).

\bibitem{pir81}
C. Piron, {\it Erkenntnis} {\bf 16}, 398 (1981).

\bibitem{pir85}
C. Piron, 'Quantum Mechanics Fifty Years After', in P. Lahti and P. Mittelstaedt, eds.,
{\it Symposium on the Foundations of Modern physics}, World scientific (1985).

\bibitem{pir90} 
C. Piron, {\it M\'ecanique Quantique Bases et Applications}, Presses Polytechniques et
Universitaires Romandes, Lausanes (1990).

\bibitem{pir96} 
C. Piron, 'Un nouveau regard sur le monde physique', preprint, Universit\'e de Gen\`eve (1996).

\bibitem{pir97}
C. Piron, 'Quantum Mechanics and Relativity: two failed revolutions', in D. Aerts, ed., {\it
Einstein Meets Magritte: White Book}, Kluwer Academic Publishers (1998).

\bibitem{qui73} 
A. Quinton, {\it The Nature of Things}, Routledge \& Kegan Paul, London and
Boston (1973).

\bibitem{rau}
H. Rauch, {\it Helv. Phys. Acta} {\bf 61}, 589 (1988).

\bibitem{red83} 
P. Redondi, {\it Galilei, eretico}, Einaudi, Turino (1983). 

\bibitem {res80} 
N. Rescher and R. Brandom, {\it The logic of inconsistency. A study in
Non-Standard Possible-World Semantics and Ontology}, Basil Blackwell, Oxford (1980).

\bibitem{rie70} 
K. Riezler, {\it Parmenides, Text, \"{U}bersetzung, Einf\"{u}hrung und
Interpretation}, Vittorio Klostermann, Frankfurt a. M. (1970).

\bibitem{ros89} 
R. Rosen, 'The Roles of Necessity in Biology', in {\it Newton to Aristotle.
Toward a Theory of Models for Living Systems}, Birkh\"auser, Boston, Basel and Berlin (1989).

\bibitem{sch85} 
E. Schr\"{o}dinger, {\it Space-Time Structure}, Cambridge University Press,
Cambridge (1950-1985). 

\bibitem{sch96} 
E. Schr\"{o}dinger, {\it Nature and the Greeks and Science and Humanism, Canto},
Cambridge University Press (1954-1996).

\bibitem {spe69} 
G. Spencer Brown,  {\it Laws of Form}, George Allen and Unwin, ltd., London (1969).

\bibitem{tou84} 
S. Toulmin, {\it Cosmopolis, the hidden agenda of modernity}, University of
Chicago Press, Chicago (1984).

\bibitem{tre96} 
H. Tredennick,  {\it Aristotle. Metaphysics}, books I-IX, Loeb XVII, Loeb
Classical Library, Harvard University Press, Cambridge, Massachusetts (1996).

\bibitem{fra}  
B.C. van Fraassen, {\it Quantum Mechanics}, Clarendon Press, Oxford (1991).

\bibitem{vla71} 
G. Vlastos, 'Reasons and causes in th Phaedo', in G. Vlastos, ed., {\it Plato: a
collection of Critical Essays}, vol I, Doubleday (1971).
 
\bibitem{vla93}
G. Vlastos, {\it Studies in Greek Philosophy, volume I: The Presocratics}, Princeton University
Press, Princeton (1993).

\bibitem{neu55}
J. von Neumann, {\it The Mathematical Foundations of Quantum Mechanics}, Princeton University
Press (1955).


\end{thebibliography}
\end{document}